# Bidding under Uncertainty: Theory and Experiments

**Amy Greenwald**
Department of Computer Science
Brown University, Box 1910
Providence, RI  02912
amy@brown.edu

**Justin Boyan**
ITA Software
141 Portland Street
Cambridge, MA  02139
jab@itasoftware.com

## Abstract

This paper describes a study of agent bidding strategies, assuming combinatorial valuations for complementary and substitutable goods, in three auction environments: sequential auctions, simultaneous auctions, and the Trading Agent Competition (TAC) Classic hotel auction design, a hybrid of sequential and simultaneous auctions. The problem of bidding in sequential auctions is formulated as an MDP, and it is argued that expected marginal utility bidding is the optimal bidding policy. The problem of bidding in simultaneous auctions is formulated as a stochastic program, and it is shown by example that marginal utility bidding is not an optimal bidding policy, even in deterministic settings. Two alternative methods of approximating a solution to this stochastic program are presented: the first method, which relies on expected values, is optimal in deterministic environments; the second method, which samples the nondeterministic environment, is asymptotically optimal as the number of samples tends to infinity. Finally, experiments with these various bidding policies are described in the TAC Classic setting.

## 1  Introduction

One of the key challenges autonomous bidding agents face is to determine how to bid on complementary and substitutable goods—i.e., goods with combinatorial valuations—in auction environments. Complementary goods are goods with superadditive valuations: $v(A\bar{B}) + v(\bar{A}B) \leq v(AB)$; substitutable goods are goods with subadditive valuations: $v(A\bar{B}) + v(\bar{A}B) \geq v(AB)$. In general, it is impossible to assign independent valuations to complementary goods, which can be worthless in isolation, or to substitutable goods, which can be worthwhile only in isolation. Thus, the simple bidding strategy "for each good $x$, bid its valuation" is inapplicable in this framework. This paper investigates a class of bidding strategies for various auction environments, assuming combinatorial valuations for complementary and substitutable goods.

Specifically, we consider three auction environments: sequential auctions, simultaneous auctions, and the hybrid of sequential and simultaneous auctions implemented in TAC Classic.[1] As the name suggests, in sequential auctions, goods are sold sequentially, in some fixed, known order. Here, agents can reason about each good in turn, basing future decisions on past outcomes. But in simultaneous auctions all goods are sold simultaneously. Here, agents must reason about all goods simultaneously, with only one opportunity to make one bidding decision that pertains to all goods. In the TAC Classic (hotel) auction design [10], auctions close sequentially, but in some random, unknown order. Here, before each auction closes, agents must reason about all goods in as yet open auctions simultaneously, but after each auction closes, agents can base future decisions on past outcomes.

Rather than attempt to reason about the valuations of goods independently, bidding agents that operate in these auction environments can reason about *marginal* valuations, or the valuation of a good $x$ relative to a set of goods $X$. In particular, if an agent holds the goods in $X$, it can ask questions such as: "what is the marginal benefit of buying $x$?" or "what is the marginal cost of selling $x$?" In doing so, the agent reasons about the *set* of goods $X \cup \{x\}$ or $X \setminus \{x\}$, relative to the set $X$—the valuations of which are well-defined. In Section 2 of this paper, we prove that the simple bidding strategy "for each good $x$, bid its (average) *marginal utility*" is optimal in sequential auctions. But not so in simultaneous auctions, as the following example shows.

---

[1]Visit http://www.sics.se/tac for details.

**Example 1.1** Consider a set of $N > 1$ goods that are being auctioned off simultaneously. Assume the value of one or more of these goods is 2, while the auction price of each good is 1, deterministically.[2] In this setting, bidding marginal utilities amounts to bidding 1 on each good. In doing so, this strategy obtains utility $2 - N < 1$. In contrast, any strategy that bids 1 on exactly one good obtains utility $2 - 1 = 1$. Thus, bidding marginal utilities is suboptimal. □

Beyond seeking an optimal bidding policy for TAC Classic hotel auctions, the goal of this research is to derive optimal solutions to the problems of bidding on goods with combinatorial valuations in sequential and simultaneous auctions. The difficulty that an agent faces in optimizing its behavior in these problems is due to the uncertainty that arises from its lack of information about the other agents' bidding strategies. In this paper, we make the simplifying assumption that the price of each good is given by an exogenous probability distribution, which is determined by the collective behavior of all competing agents, but which ignores the behavior of the optimizing agent. Consequently, even if the optimizing agent places a winning bid, the price of the good does not depend on this bid.

The bidding strategies analyzed in this study, which were inspired by agent strategies in the international Trading Agent Competition [10] (TAC Classic), are all based on the notion of marginal utility. Specifically, we study expected marginal utility bidding, implemented in ATTAC-01 [9], and variants of policy search, implemented in ROXYBOT-00 [7] and ROXYBOT-02 [6]. Note that these teams were two of the few to exploit stochastic price information in their agent design, beyond computing straightforward expected values [11]. Although marginal utility bidding is not optimal in simultaneous auctions, we prove that it is the optimal bidding policy in sequential auctions, and we show empirically that it is a reasonable heuristic for bidding in TAC Classic hotel auctions.

## 2 Sequential Auctions

In this section, we formulate the bidding problem in sequential auctions—a sequential decision problem—as a Markov decision process (MDP), and we compute the optimal policy. As an example, consider the MDP depicted in Figure 1, which represents an agent's bidding problem in two sequential auctions for two goods $A$ and $B$. There are seven states in this MDP, drawn in three stages. In the first stage, a bidding decision is made regarding good $A$, and in the second stage, a bidding decision is made regarding good $B$. Along the way, the agent earns negative rewards equal to the price of each good for which it places a winning bid. A winning bid for the $j$th good is a bid for an amount that is greater than or equal to the (uncertain) price of that good. Transitions are stochastic: *e.g.*, given bid $b'$, the system transitions from state $(\{A\}, 1)$ to state $(\{A, B\}, 2)$ with probability equal to the probability that $b'$ is a winning bid; otherwise the system transitions to state $(\{A\}, 2)$. In the final stage, the agent earns rewards equal to the valuation of the subset of the set $\{A, B\}$ it successfully acquires.

---
[2]Think of the BUY IT NOW option at eBay.com, which serves as an example of deterministic auction prices.

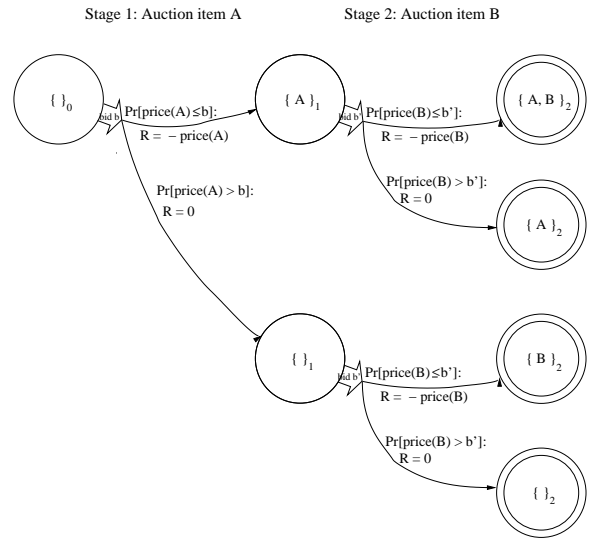

Figure 1: The sequential auction problem as an MDP: An example with two goods $A$ and $B$.

In general, a finite set of $n$ goods $N = \{x_1, \ldots, x_n\}$ is given. Let $J \subseteq N = \{x_1, \ldots, x_j\}$. Now a state at stage $j$ is denoted $(X, j)$, where $X \subseteq J$ is the set of current holdings and good $x_{j+1}$ is up for auction. The actions available at each state are real-valued bids. (Note that the action space is continuous.) A bid $b$ for good $x_j$ at state $(X, j-1)$ is declared to be a winning bid if the good's randomly sampled price $p_j \leq b$. All other bids are losing bids. Transitions, which depend on state-action pairs, are stochastic. Given bid $b$, a transition is made from state $(X, j-1)$ to state $(X \cup \{x_j\}, j)$ with probability equal to the probability that $b$ is a winning bid; otherwise a transition is made from $(X, j-1)$ to $(X, j)$. Rewards are associated with state-action-state triples at all stages $0 \leq j \leq n-1$ as follows: a winning bid at stage $j$ earns reward $-p_j$, where $p_j$ is the price of the $j$th good; a losing bid earns no rewards. In addition, a valuation function $v : 2^N \to \mathbb{R}$, which assigns a valuation to each subset of $N$, defines the rewards at stage $n$, at which point the auction ends.

**Definition 2.1** The bidding problem in sequential auctions is defined by the following MDP:

- States: each state at stage $j$, for $0 \leq j \leq n$, is denoted $(X, j)$, where $X \subseteq J$.

- Actions: the actions available at all states $(X, j-1)$ at stage $j-1$ are bids $b_j \in \mathbb{R}^+$ on good $x_j$.

- Transitions: $P((X \cup \{x_j\}, j)|(X, j-1), b) = F_j(b)$, and $P((X, j)|(X, j-1), b) = 1 - F_j(b)$, where $F_j(x) = \Pr[p_j \leq x]$ for good $j$: i.e., $F_j$ is the *cdf* over prices $p_j$.

- Rewards:
  - For all states $(X, n)$ with $X \subseteq 2^N$ at stage $n$, $R((X, n)) = v(X)$.
  - For all other states $(X, j)$ with $X \subseteq J$ at stage $j$, if $b$ is a winning bid for good $j$, then reward $-p_j$ is incurred; otherwise, no cost is incurred. Define the following:

$$r((X, j-1), b, p) = \begin{cases} -p & \text{if } p \leq b \\ 0 & \text{otherwise} \end{cases} \quad (1)$$

Now

$$R((X, j-1), b) = \int_{-\infty}^{\infty} r((X, j-1), b, p) f_j(p)\, dp \quad (2)$$

where $f_j = F_j'$ is the *pdf* over prices $p_j$.

The optimal policy in this MDP is described by Bellman's equations [1]:

$$\pi((X, j)) \in \arg\max_b Q((X, j), b) \quad (3)$$

$$Q((X, j), b) = R((X, j), b) + F_j(b)V((X \cup \{x_{j+1}\}), j+1) + (1 - F_j(b))V((X, j+1)) \quad (4)$$

$$V((X, j)) = \max_b Q((X, j), b) \quad (5)$$

**Theorem 2.2** *The following bidding policy is optimal in this MDP: at state $((X, j-1))$, place bid $b_j^* = V((X \cup \{x_j\}), j) - V((X, j))$.*

**Proof 2.2** For arbitrary $p_j$, two cases arise: Case $b_j^* \geq p_j$: If $b \geq p_j$, then $Q((X, j-1), b) = V(((X \cup \{x_j\}), j)) - p_j = Q((X, j-1), b_j^*)$. If, however, $b < p_j$, then $Q((X, j-1), b) = V((X, j)) < V(((X \cup \{x_j\}), j)) - p_j = Q((X, j-1), b_j^*)$, since $b_j^* > p_j$, by assumption. The case in which $b_j^* < p_j$ is symmetric. □

This MDP formulation is related to that of Boutilier, *et al.* [4]. The most notable difference is in the two state representations. In their model, states consist of current holdings together with an endowment, which decreases as goods are acquired. In our model, rather than decrease an endowment, expenses are modeled as negative rewards. Our representation leads to a notable savings in the size of the state space, but cannot model an agent with a finite budget. Note that Theorem 2.2 is not applicable in their model.

### 2.1 Marginal Utility

In this section, we present an interpretation of the optimal bidding policy in sequential auctions in terms of marginal utilities. Computing marginal utilities depends on solving the so-called *acquisition problem* [5]: "Given the set of goods that I already own, and given market prices and supply, on what set of additional goods should I place bids so as to maximize my utility: i.e., valuation less costs?"

**Definition 2.3** Given a set of goods $X$ that an agent already owns, a set of goods $Y$ supplied by the market s.t. $X \cap Y = \{\}$, a vector of prices $\vec{p}$ with $p_k \in \mathbb{R}_+$ for all $k \in Y$, and a combinatorial valuation function $v : 2^{X \cup Y} \to \mathbb{R}$ satisfying free disposal (i.e., if $Y \subseteq X$, $v(Y) \leq v(X)$),

- the *acquisition* function $\alpha(X, Y, \vec{p})$ is defined as follows:

$$\alpha(X, Y, \vec{p}) = \max_{Z \subseteq X \cup Y} \left( v(Z) - \sum_{k \in Z \cap Y} p_k \right) \quad (6)$$

- the *marginal utility* of good $x \notin X \cup Y$ is defined as follows: $\mu(x, X, Y, \vec{p}) = \alpha(X \cup \{x\}, Y, \vec{p}) - \alpha(X, Y, \vec{p})$.

In words, the marginal utility of good $x \notin X \cup Y$ is the difference between the utility of $X \cup \{x\} \cup Y$, assuming $x$ costs 0, and the utility of $X \cup Y$ (equivalently, the utility of $X \cup \{x\} \cup Y$, assuming $x$ costs $\infty$).

**Definition 2.4** Given a set of goods $X$ that an agent already owns, a set of goods $Y$ supplied by the market s.t. $X \cap Y = \{\}$, a joint probability density function $f$ describing the prices of the goods in $Y$, and a combinatorial valuation function $v : 2^{X \cup Y} \to \mathbb{R}$,

- the *expected* acquisition function $\overline{\alpha}(X, Y)$ is defined as follows:

$$\overline{\alpha}(X, Y) = \mathbb{E}_{\vec{p}}[\alpha(X, Y, \vec{p})] = \int_{\vec{p}} \alpha(X, Y, \vec{p}) f(\vec{p}) d\vec{p} \quad (7)$$

- the *expected* marginal utility of good $x \notin X \cup Y$ is given by: $\overline{\mu}(x, X, Y) = \overline{\alpha}(X \cup \{x\}, Y) - \overline{\alpha}(X, Y)$.

Thus, the expected marginal utility at state $(X, j-1)$ is the difference between the expected value of the solution to acquisition assuming $x_j$ costs 0, and the expected value of this solution assuming $x_j$ costs $\infty$.

The proof of the following (key) lemma is omitted.

**Lemma 2.5** *For all states $(X, j)$ in the MDP, $V((X, j)) = \overline{\alpha}(X, Y_j)$, where $Y_j = \{x_{j+1}, \ldots, x_n\}$.*

**Corollary 2.6** *For all states $(X, j)$ in the MDP, and for all goods $x_j$, $\overline{\mu}(x_j, X, Y_j) = V((X \cup \{x_j\}, j)) - V((X, j))$, where $Y_j = \{x_{j+1}, \ldots, x_n\}$.*

**Proof 2.7** The proof follows immediately from Lemma 2.5:

$$\begin{aligned}\overline{\mu}(x_j, X, Y_j) &= \overline{\alpha}(X \cup \{x_j\}, Y_j) - \overline{\alpha}(X, Y_j) \\ &= V(((X \cup \{x_j\}), j)) - V((X, j))\end{aligned}$$

**Corollary 2.8** *Expected marginal utility bidding is the optimal bidding policy in sequential auctions.*

**Proof 2.8** The proof follows from Theorem 2.2 together with Corollary 2.6. □

Returning to the example in Figure 1, the optimal policy at state $(\{\}, 0)$ is simply to bid the expected marginal utility of good $A$, which is the difference between the value of state $(\{A\}, 1)$ and state $(\{\}, 1)$. Similarly, the optimal policy at state $(\{A\}, 1)$ is to bid the expected marginal utility of good $B$ given that the agent already owns good $A$, which is the difference between the value of states $(\{A, B\}, 2)$ and $(\{A\}, 2)$; and, the optimal policy at state $(\{\}, 1)$ is to bid the expected marginal utility of good $B$ given that the agent owns nothing as of yet, which is the difference between the values of states $(\{B\}, 2)$ and $(\{\}, 2)$. In what follows, we abbreviate expected marginal utility bidding by $\overline{\text{MU}}$, and we abbreviate marginal utility bidding (assuming deterministic prices) by MU.

## 3 Simultaneous Auctions

In simultaneous auctions, an agent must make all its bidding decisions *a priori*, with only probabilistic knowledge of what it may or may not win.

For example, suppose that a camera and a flash are being auctioned off simultaneously, and that an agent assigns valuation $750 to these two goods together, but assigns valuation $0 to either good alone. In addition, suppose that the price of the flash is known with certainty: it is $50; but the price of camera will be $500 with probability $\frac{1}{2}$ and $1000 with probability $\frac{1}{2}$. Given these assumptions, what is an optimal bidding policy? If it happens that the camera sells for $500, then it is optimal to bid $500 for the camera and $50 for the flash. But, if it happens that the camera sells for $1000, then it is optimal to bid $0 for both the camera and the flash. Evaluating these two policies, bidding ($0,$0) yields $0 utility, while bidding ($500,$50) yields $200 utility half the time, and $-$50 utility half the time. Thus, the expected value of bidding ($500,$50) is $75, and this is the best possible.

In this section, we formulate the problem of bidding in simultaneous auctions as a stochastic program whose solution is an optimal bidding policy in the expected sense. Here, the expectation is computed over all possible stochastic outcomes (*a.k.a.* scenarios): e.g., if there are two goods, this set of scenarios includes win good 1, win good 2, win both goods, win neither good. Since the number of scenarios is exponential in the number of goods, computing an optimal solution to this stochastic program is intractable for large numbers of goods. We discuss three methods for approximating an optimal bidding policy in this environment: one heuristic approach (expected MU bidding), and two approximation schemes (the so-called *expected value method* with MU bidding and a stochastic sampling technique). In the next section, we describe experiments with all three of these strategies in the TAC Classic auction framework.

### 3.1 Problem Statement

Given valuation function $v : 2^X \to \mathbb{R}$, let $\vec{v}$ denote the vector of valuations, with $v_i$ as the valuation of the $i$th subset of $X$. As in Section 2.1, good prices are described by the joint probability function $f$. We seek a set of bids that maximizes total expected utility: i.e., expected valuation less expected cost. That is, we seek a set of bids *today*, before any uncertainty is resolved, that maximizes our expected utility *tomorrow*, after all uncertainty is resolved. Before tackling the bidding problem, we first describe the *allocation* problem, which arises after all uncertainty is resolved, since it is at this point that all winnings are allocated to bundles.

Let $\vec{p}$ denote the vector of prices with $p_{jk} \in \mathbb{R}_+$ as the price of the $k$th copy of good $j$. Let the continuous decision variables $b_{jk} \in \mathbb{R}_+$ denote the bid placed on the $k$th copy of good $j$; let the binary decision variables $a_{ijk} \in \{0, 1\}$ indicate whether or not the $k$th copy of good $j$ is allocated to bundle $i$. Define the following:

$$\pi(\vec{a}, \vec{b}, \vec{p}, \vec{v}) = -\sum_{jk} p_{jk}(\mathbf{1}[p_{jk} \leq b_{jk}])$$
$$+ \sum_i v_i \left( \prod_{j \in i} \mathbf{1}\left[ n_{ij} \leq \sum_k a_{ijk} \mathbf{1}[p_{jk} \leq b_{jk}] \right] \right) \quad (8)$$

where $n_{ij}$ denotes the number of copies of good $j$ that are essential to bundle $i$, and $\mathbf{1}[x \leq y]$ is an indicator function that evaluates to 1 if $x \leq y$ and otherwise evaluates to 0. According to $\pi$, goods for which winning bids are placed incur costs, but are also allocated to bundles that secure valuations, as long as enough copies of all goods that are essential to a bundle are indeed allocated to that bundle. The *allocation* problem can be described by the following integer program:

$$\max_{\vec{a}} \pi(\vec{a}, \vec{b}, \vec{p}, \vec{v}) \tag{9}$$

$$\text{subject to:} \quad \sum_i a_{ijk} \leq 1, \quad \forall j, k \tag{10}$$

$$a_{ijk} \in \{0, 1\}, \quad \forall i, j, k \tag{11}$$

The first set of constraints states that each copy $k$ of each good $j$ can be allocated to at most one bundle.

The following stochastic program [3] solves the bidding problem in simultaneous auctions:

$$\max_{\vec{b}} \int_{\vec{p}} \max_{\vec{a}} \pi(\vec{a}, \vec{b}, \vec{p}, \vec{v}) f(\vec{p}) d\vec{p} \tag{12}$$

$$\text{subject to:} \quad b_{jk} \in \mathbb{R}_+, \quad \forall j, k \tag{13}$$

Notice that the allocation problem is nested inside the bidding problem.

Later, we refer to the deterministic version of the bidding problem as *completion*: given prices $\vec{p}$,

$$\max_{\vec{a}, \vec{b}'} \pi'(\vec{a}, \vec{b}', \vec{p}, \vec{v}) \tag{14}$$

$$\text{subject to:} \quad b'_{jk} \in \{0, 1\}, \quad \forall j, k \tag{15}$$

where

$$\pi'(\vec{a}, \vec{b}', \vec{p}, \vec{v}) =$$

$$\sum_i v_i \left( \prod_{j \in i} \mathbf{1} \left[ n_{ij} \leq \sum_k a_{ijk} b'_{jk} \right] \right) - \sum_{jk} p_{jk} b'_{jk} \tag{16}$$

### 3.2 Heuristics & Approximation Algorithms

In this section, we discuss three methods for approximating an optimal solution to this stochastic program: one heuristic approach inspired by ATTAC-01 (expected MU bidding), and two approximation schemes inspired by ROXYBOT (the so-called *expected value method* with MU bidding and a stochastic sampling technique). First, we show that the heuristic of bidding expected marginal utilities, while optimal in sequential auctions, is suboptimal in simultaneous auctions. Second, we show that the expected value method with MU bidding is also suboptimal, although this approach is optimal when prices are deterministic. Third, we discuss an asymptotically optimal sampling method: i.e., as the number of samples grows, the value of the approximate solution approaches the value of the stochastic programming solution.

#### 3.2.1 ATTAC-01: Expected MU Bidding

In Example 1.1, in the introduction, we argued that marginal utility bidding is suboptimal in simultaneous auctions with deterministic prices. We present a second example here in which prices are nondeterministic and *expected* marginal utility bidding is suboptimal.

**Example 3.1** Let $v(x) = v(y) = v(xy) = 1$. Assume the prices of goods $x$ and $y$ are described by the following bipolar distribution: $p(a) = 1$, with probability $\frac{1}{2}$, and $p(a) = 101$, with probability $\frac{1}{2}$, for all $a \in \{x, y\}$. Now expected marginal utility bidding gives rise to the policy "Bid 1 on both goods" in this example, since $\mu(x, \emptyset, \{y\}, \vec{p}) = \mu(y, \emptyset, \{x\}, \vec{p}) = 1$ under all price samples $\vec{p}$. But then expected marginal utility bidding earns expected utility $-\frac{1}{4}$. The policy "Bid 0 on both goods" (which earns expected utility 0) dominates expected marginal utility bidding in this example.

| $x$ | $y$ | $\mu(x)$ | $\mu(y)$ | Evaluation |
|---|---|---|---|---|
| 1 | 1 | 1 | 1 | -1 |
| 1 | 101 | 1 | 1 | 0 |
| 101 | 1 | 1 | 1 | 0 |
| 101 | 101 | 1 | 1 | 0 |
| Average | | 1 | 1 | $-\frac{1}{4}$ |

#### 3.2.2 ROXYBOT-00: Expected Value Method with MU Bidding

In this section, we show (i) the expected value method with MU bidding (EVMU) is optimal when prices are deterministic; but, (ii) in general, the expected value method, and therefore EVMU, is suboptimal.

It is common to approximate the solution to stochastic programs using the so-called *expected value method* (see, for example, [3]). This method solves the deterministic version of the problem, assuming all stochastic inputs have deterministic values equal to their expected values. Applying this method to our stochastic program yields a solution to Equation 14: i.e., an optimal set of goods on which to bid. As in the deterministic setting, it is straightforward to transform a solution to this problem into an optimal bidding policy: bid $\$\infty$, whenever $b_j = 1$; bid $\$0$, whenever $b_j = 0$. But, in general the expected value method is suboptimal.

Recall the discussion of the camera and the flash introduced at the beginning of this section. One attempt to approximate the optimal solution (in the expected sense) is obtained by solving the deterministic variant of the problem: assume the price of the camera is $\$750$ (its expected price), while the price of the flash is $\$50$. Under these assumptions, the cost exceeds the valuation of the camera and the flash; thus, the optimal policy is to bid ($\$0$,$\$0$), which yields $\$0$ utility.

In this example, the so-called *value of stochastic information* (i.e., the difference between the solutions to Equation 12 and Equation 14) is $75.

The following example shows that there is value to stochastic information not only in simultaneous auctions, but even in an auction for only one good (which is both a sequential and a simultaneous auction).

**Example 3.2** Consider only one good $a$ of value $100. Suppose $a$'s price is $1 with probability .9, but that $a$'s price is $1 million with probability .1. Thus, the expected price of good $a$ is roughly $100,0010. The optimal policy using the expected value method is to bid $0, which scores $0. But now consider the bidding policy "bid $100." This policy scores $99 with probability .9, and $0 with probability .1. Thus, on average, this policy scores roughly $89. "Bid $100" dominates the expected value method in this example. Indeed, "bid $100," which corresponds to bidding expected marginal utility, is optimal, since this auction is sequential. □

In the introduction, we made an important simplifying assumption, namely, the price of each good is given by an exogenous probability distribution, which is determined by the collective behavior of all competing agents, but which ignores the behavior of the optimizing agent. In TAC Classic hotel auctions, for example, this assumption is violated: an agent's bid *can* impact the prices of goods—a winning agent could even pay what it bids. As a heuristic that is applicable in this more general setting, we propose the following bidding policy: bid marginal utilities on all goods in an optimal set $A^*$ computed using the expected value method. If it is possible that an agent could pay what it bids, then marginal utility seems to be a reasonable upper bound on what it should bid, since bidding marginal utility on some good $x \in A^*$ is indeed optimal if the prices of all other goods $y \neq x \in A^*$ are deterministic. In summary, we propose the following bidding policy: (i) set $p_j$ equal to the mean of $P_j$, (ii) solve the completion problem (i.e., Equation 14), and (iii) bid marginal utilities on all goods in the optimal completion: i.e., all goods for which $b_j = 1$. We call this strategy EVMU. It was implemented in ROXYBOT-00 [7], and it was shown that *EVMU bidding is optimal in simultaneous auctions when prices are deterministic* in Greenwald [6].

### 3.2.3 ROXYBOT-02: MU Bidding Policy Search

As a third means of approximating an optimal solution to the problem of bidding in simultaneous auctions, one possible technique called *Sample Average Approximation* (SAA) method solves the stochastic program using only a subset of the scenarios, randomly sampled according to the scenario distribution. An important theoretical justification for this method is that as the sample size increases, the solution converges to an optimal solution in the expected sense. Indeed, the convergence rate is exponentially fast [8]. In Benisch, *et al.* [2], we apply this technique to the TAC SCM scheduling problem.[3]

Here we explore an alternative means of approximating an optimal solution to the stochastic program, namely policy search, which in the absence of any clever heuristics, is simply brute-force, generate-and-test. This solution technique generates a set of candidate policies, evaluates them, and selects the best one. Candidates are evaluated over multiple samples: for each sample, the candidate's score is computed, and scores are averaged over all samples (as in the policy evaluation process in Example 3.1). If it were to evaluate all possible candidate policies under infinitely many samples, this method would tend towards outputting an optimal bidding policy.

A variant of this approach is at the heart of ROXYBOT-02, which generates candidate policies via the following heuristic: (i) determine $p_j$ by sampling from the price distributions; (ii) solve the completion problem (i.e., Equation 14), and (iii) bid marginal utilities on all goods in the optimal completion: i.e., all goods for which $b_j = 1$. More generally, ROXYBOT-02 can generate policies according to any of the aforementioned algorithms: EVMU, MU, or expected MU. By including in the space of candidate policies those generated by these alternative strategies, we can ensure (probabilistically) that ROXYBOT-02 dominates the others.

In TAC-02, ROXYBOT-02 generated its candidate policies using ROXYBOT-00's internals. In Example 3.2, this instantiation of ROXYBOT-02 generates two policies: if the sample price is $1, its bidding policy is "bid $100;" if the sample price is $1 million, its bidding policy is "bid $0." "Bid $100" scores $89, on average, whereas "bid $0" scores $0. Thus, ROXYBOT-02 employs the policy "bid $100", and scores $89, on average. This form of policy search outperforms EVMU in this example. Indeed, there is value in exploiting stochastic information beyond expected values. In the next section, we demonstrate this effect in TAC Classic.

## 4 Experiments

Our analysis of agent bidding strategies in the previous two sections was based on the assumption that prices are determined exogenously. In particular, MU is optimal in sequential auctions if prices are deterministic and exogenous; EVMU is optimal in simultaneous auctions if prices are deterministic and exogenous; and, policy search is approximately optimal in simultaneous auctions, even if prices are uncertain, but still exogenous. In this section, we discuss experiments designed to ascertain the power of these strategies in the TAC Classic hotel auctions, a hybrid of sequential and simultaneous auctions, in which prices are endogenous, rather than exogenous.

---

[3]See www.sics.se/tac for a description of TAC SCM.

In TAC Classic hotel auctions, the TAC seller auctions off 16 hotel rooms in ascending, multi-unit, sixteenth price auctions. These auctions close sequentially. The order of the auction closings is unknown to the agents. In fact, MU bidding is optimal even in sequential auctions which close in some random, unknown order *if bids can be withdrawn*. The difficulty in TAC classic hotel auctions is that bids cannot be retracted. Moreover, when agents submit bids, they must "beat the quote," according to the following rules:

> Let $a$ be the the sixteenth highest price. Any new bid $b$ must satisfy the following conditions to be admitted to the auction: (i) $b$ must offer to buy at least one unit at a price of $a + 1$ or greater; (ii) if the agent's current bid $c$ would have resulted in a purchase of $q$ units, then the new bid $b$ must offer to buy at least $q$ units, again at a price of $a + 1$ or greater.

### 4.1 Setup

In our experiments, we pitted 4 TAC agents bidding according to one strategy against 4 TAC agents bidding according to another strategy (*e.g.*, 4 RoxyBot-02 agents vs. 4 RoxyBot-00 agents). We refer to each set of 4 TAC agents in one game as a team. We played numerous games between pairs of teams—exact numbers depended on which teams were participating. In RoxyBot-02, we arbitrarily fixed the number of samples $n = 50$. No attempt was made to optimize this parameter. None of the other algorithms used in this study—RoxyBot-00, MU, and $\overline{\text{MU}}$—have any tunable parameters.

Before running any experiments, we played 500 training games between RoxyBot-00 and MU, initializing price estimates to the competitive equilibrium prices derived in Wellman, *et al.* [11]. Using data collected from these training games, we generated distributions over clearing prices for each good. The distributions were represented by a lookup table over five salient features of the domain, the details of which are beyond the scope of this paper. In each cell of the table, 10 numbers were stored, corresponding to the predictions at percentiles $5, 10, 15, \ldots, 95$. This representation was chosen for its simplicity and its weak assumptions about the shape of the underlying price distributions. We sought to capture the highly skewed and multimodal distributions that arise in practice in TAC games.

A sample output after the run of one game instance is shown in the table below.

| Agent | Score | Rank |
|-------|-------|------|
| RoxyBot-02 | 3802 | 1 |
| RoxyBot-02 | 3116 | 6 |
| RoxyBot-02 | 3166 | 5 |
| RoxyBot-02 | 3447 | 4 |
| RoxyBot-00 | 2521 | 8 |
| RoxyBot-00 | 3696 | 2 |
| RoxyBot-00 | 2788 | 7 |
| RoxyBot-00 | 3600 | 3 |

Generally speaking, scores in our experiments are not as high as scores in actual competitions, because bidding marginal utilities, as do all eight agents in our experiments, tends to lead to high prices.

### 4.2 Evaluation

To evaluate our results, we use two statistical tests: the $z$-test, which we use to compare scores, and the Wilcoxon test, which we use to compare rankings. In our context, the inputs to the $z$-test are two sample datasets of scores, one per team, over many game instances. The $z$-test outputs the probability that the difference between the means of these datasets is positive: *i.e.*, the probability that the mean of the second is greater than the mean of the first. Our input to the Wilcoxon test[4] is a list of pairs of average rankings, one per game. The test measures the significance of the difference between these rankings.

### 4.3 Results

Our experimental results are depicted numerically in Table 1 and graphically in Figure 2. MU outperforms expected MU; RoxyBot-00 outperforms MU; and, RoxyBot-02 outperforms RoxyBot-00. Moreover, these results are transitive: RoxyBot-00 outperforms expected MU bidding; RoxyBot-02 outperforms MU bidding; and RoxyBot-02 outperforms expected MU bidding.

The numbers in Table 1 reveal that with high confidence, RoxyBot-02 is expected to score higher than RoxyBot-00 and MU. On the other hand, although RoxyBot-00 is expected to score higher than MU, the lower confidence level supporting this conclusion makes this result less credible. The outcome of all

---

[4]For a description of the Wilcoxon test, visit http://fonsg3.let.uva.nl/Service/Statistics/Signed_Rank_Test.html.

| Teams | Means | | $z$-test | Wilcoxon | Games |
|---|---|---|---|---|---|
| $\overline{\mathrm{MU}}$ < MU | 964 | 1908 | .999 | .999 | 25 |
| MU < RoxyBot-00 | 1508 | 1612 | .793 | .803 | 75 |
| RoxyBot-00 < RoxyBot-02 | 1837 | 2031 | .977 | .996 | 50 |
| $\overline{\mathrm{MU}}$ < RoxyBot-00 | 1334 | 2034 | .999 | .999 | 25 |
| MU < RoxyBot-02 | 1705 | 1987 | .976 | .993 | 50 |
| $\overline{\mathrm{MU}}$ < RoxyBot-02 | 915 | 1920 | .999 | .999 | 25 |

Table 1: Numerical Results: Means, $z$-test, Wilcoxon test, and Sample Size.

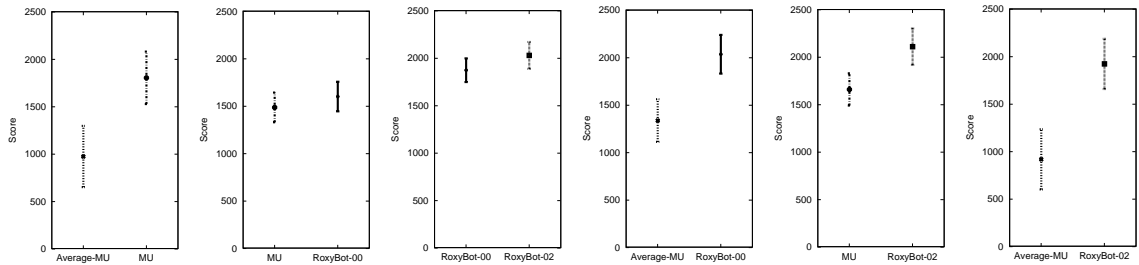

Figure 2: Graphical Results: 95% Confidence Intervals on the Means. All agent strategies clearly outperform expected MU bidding. Other graphs reveal narrower performance distinctions.

Wilcoxon tests, and the graphs depicted in Figure 2, reinforce the outcome of the $z$-tests.

Notably, with high confidence in the $z$-tests and Wilcoxon tests, all strategies are expected to score higher than expected MU, and with 95% confidence, all strategies outperform expected MU. Since expected MU bidding is the optimal policy in sequential auctions, we conclude that TAC classic hotel auctions are more similar in spirit to simultaneous auctions than to sequential auctions.